\documentclass[printer]{aa}

\bibpunct{(}{)}{;}{a}{}{,}

\usepackage{graphicx}
\usepackage[varg]{txfonts}
\def\ks{km s$^{-1}$}
\def\d{$^\circ$}

\def\s{$^{\prime\prime}$}

\def\cm3{cm$^{-3}$}

\def\2{$^{12}$CO}
\def\3{$^{13}$CO}
\def\msol{M$_\odot$}

\def\cm2{cm$^{-2}$}
\begin{document}

\title{Near-IR imaging towards a puzzling YSO precessing jet}
\author {S. Paron \inst{1,2}
\and C. Fari\~{n}a \inst{3}
\and M. E. Ortega \inst{1}
}

\institute{Instituto de Astronom\'\i a y F\'\i sica del Espacio (IAFE),
             CC 67, Suc. 28, 1428 Buenos Aires, Argentina\\
             \email{sparon@iafe.uba.ar, mortega@iafe.uba.ar}
\and CBC and FADU - Universidad de Buenos Aires, Ciudad Universitaria, Buenos Aires, Argentina
\and Isaac Newton Group of Telescopes, E-38700, La Palma, Spain \\
              \email{cf@ing.iac.es}
}

\offprints{S. Paron}

   \date{Received <date>; Accepted <date>}

\abstract{}{The study of jets related to stellar objects in formation is important because it enables us to understand the history of how
the stars have built up their mass. At present there are many studies concerning jets towards low-mass young stellar objects, while
equivalent studies towards massive or intermediate-mass young stellar objects are scarce. 
In a previous study, based on $^{12}$CO J=3--2 and public near-IR data, 
we found highly misaligned molecular outflows towards the infrared point source UGPS J185808.46+010041.8 (IRS along this work), 
and some infrared features suggesting the existence of a precessing jet.}
{Using near-infrared data acquired with Gemini-NIRI at the {\it JHKs}-broad-bands and narrow-bands centered at the emission lines of 
[FeII], H$_{2}$ 1-0 S(1), H$_{2}$ 2-1 S(1), Br$\gamma$, and CO 2-0 (bh), we studied the circumstellar environment of IRS with an 
angular resolution between 0\farcs35 and 0\farcs45.}
{The emission in the {\it JHKs}-broad-bands shows, with great detail, the presence of a cone-like shape nebula extending to the north/northeast 
of the point source, which appears to be attached to it by a jet-like structure. In the three bands the nebula is resolved in a twisted-shaped 
feature composed by two arc-like features and a bow shock-like structure seen mainly in the {\it Ks}-band, which strongly suggests the presence
of a precessing jet. An analysis of proper motions based on our Gemini observations and UKIDSS data gives additional support to the 
precession scenario. We are presenting one of the best resolved cone-like nebula likely related to a precessing jet up to date. 
The analysis of the observed near-infrared lines shows that the H$_{2}$ is collisionally excited,
and the spatially coincidence of the [FeII] and H$_{2}$ emissions in the closer arc-like feature
suggests that this region is affected by a J-shock.  The second arc-like feature presents H$_{2}$ emission without
[FeII] which suggests the presence of a nondisociative C-shock or a less energetic J-shock. The H$_{2}$ 1-0 S(1) continuum subtracted image, 
reveals several knots and filaments at a larger spatial scale around IRS, in perfect matching with the distribution
of the red and blueshifted molecular outflows discovered in our previous work. 
A not resolved system of YSOs is suggested to explain the distribution of the analyzed near-infrared features 
and the molecular outflows, which in turns explain the jet precession through tidal interactions.
    }{}

\titlerunning{Near-IR imaging of a YSO precessing jet}
\authorrunning{S. Paron et al.}


\maketitle

\section{Introduction}

From observations and theoretical studies it is known that when a star forms, jets arise from a region in close proximity to the 
accreting source, which extract mass and angular momentum from the underlying disk.
The study of jets related to stellar objects in formation enables us to understand the history of how
the stars have built up their mass. As the jets penetrate their surroundings, 
they transfer momentum and accelerate matter, generating the ubiquitous molecular outflows commonly found in the surroundings of
young stellar objects (YSOs) \citep{reip01,mckee07}. Outflow (a)symmetries provide information about the dynamical environment 
of the engine and the interstellar medium in which they spread; the so called S- and Z-shaped symmetries indicate that the outflow 
axis has changed over time,
probably due to precession induced by a companion, or interactions with sibling stars in a cluster \citep{bally07},
while C-shaped bends may indicate motion of surrounding gas or motion of the outflow source itself.
More recently, it has been suggested that the jet precession may be produced also by the misalignment between the protostar rotation
axis and the magnetic fields \citep{ciardi10,lewis15}.
At present there are many studies concerning the (a)symmetries of jets and outflows, mainly towards the Orion and Carina Nebula, which are very
rich in HH objects related to low-mass YSOs (e.g. \citealt{leflo07,bally09,bally12,davis11,reiter15}).
However, equivalent studies towards massive or intermediate-mass YSOs are less common (e.g. \citealt{prei03,weigelt06,paron13}).  
Factors as the complexity of the environments, stellar multiplicity, and scarceness of massive YSOs in our proximity, 
make the observational studies towards these objects both, challenging and encouraging.

\begin{figure*}[tt!]
\centering
\includegraphics[width=18cm]{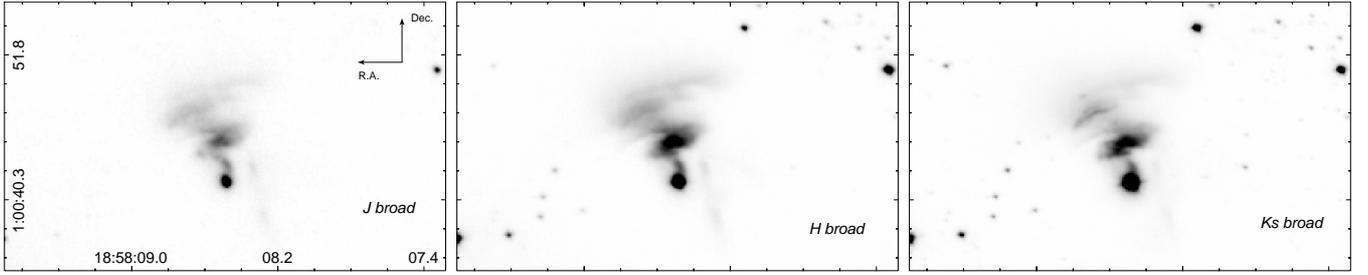}
\caption{{\it JHKs} broad-bands emission.}
\label{jhkbroad}
\end{figure*}

In a previous paper (\citealt{paron14}; hereafter Paper\,I) we presented results from the observation of highly misaligned molecular outflows 
towards the infrared (IR) point source UGPS J185808.46+010041.8 \citep{lucas08,lucas12}. Analyzing public UKIDSS near-IR data ({\it JHKs} 
broad-bands) 
extracted from the WFCAM Science Archive we found some diffuse emission showing a cone-like nebula  
related to the point source, which could be due to a cavity cleared in the circumstellar material by a precessing jet. 
This source, located at a distance of about 1.1 kpc \citep{lumsden13}, was suggested to be a young intermediate-mass protostar
(about 3 \msol) from an spectral energy distribution (SED) analysis (Paper\,I).

In order to study the origins of the misalignment of the molecular outflows and the possibility of a precessing jet, 
we obtained high-resolution images of this source with NIRI, at the Gemini Telescope, using a set of broad- and narrow-band near-IR filters.

\section{Observations and data reduction}

In this study we analyze several near-IR broad- and narrow-band images (see Table 1) towards the source UGPS J185808.46+010041.8 (hereafter IRS). 
The images were acquired with NIRI, Near InfraRed Imager and Spectrometer \citep{hodapp03} at Gemini-North 8.2-m telescope. 
The observations were carried out during August and October 2014, and April 2015 in queue mode (Band-1 Program GN-2014B-Q-35). 
NIRI was used with the f/6 camera that provides a plate scale of 0\farcs117 pix$^{-1}$ in a field of view of 120\s$\times$120\s. 

To optimize telescope time the images were performed following a dither pattern in which the offsets directions and amplitudes were selected
to be able to perform the near-IR background correction using the on-source images. For the dither pattern, particular careful was 
taken to avoid the contamination on the IRS source and its associated nebular emission by a saturated field star,
located at about 30 arcsec from the IRS source, which residuals inevitably remain in subsequent images.

For data reduction NIRI images were firstly passed through {\it nirlin.py}, a Python script provided by the Gemini Observatory that applies a per-pixel 
linearity correction. All subsequent processes as background correction, images co-addition, and astrometric
solution were performed with Theli \citep{schi13,erben05}. 
For the absolute astrometric solution, stars in the field from the 2MASS 6X Point Source Working Database Catalog \citep{cutri} were used.
 
Table\,\ref{obs} lists the filters used with their central wavelength and width, the effective spatial resolution measured as the average FWHM 
of point sources in the final co-added images, the number of individual frames, and the effective exposure times of the final 
co-added images. All images were later normalized to 1 sec.

The {\it H}-cont image was used to subtract the continuum of [FeII] while the {\it K}-cont one to subtract the continuum of   
H$_{2}$ 1-0 S(1), Br$\gamma$, H$_{2}$ 2-1 S(1), and CO 2-0 (bh). For the continuum subtraction the images were convolved with an elliptical 
Gaussian function using the {\it gauss} IRAF\footnote{IRAF is distributed by the NOAO, which are operated by the AURA, Inc., 
under cooperative agreement with the NSF.} task to achieve a similar PSF for the point sources at the central area of both images. 
In this process the effective resolution of both images was degraded to a similar value. After convolution the images were scaled to account 
for the differences in filter width and throughput, and other effects derived from observing conditions in different nights 
(both instrumental and environmental). An initial scale factor was derived from aperture photometry of point sources in the central part of 
the field, this value was checked to be in agreement with the values derived from the ratio of filters transmissions.
The initial scale factors were later fine-tuned by visual inspection of the residuals in subtracted images.
The object nebular emission in all the subtracted images is above 5$\sigma$ over the background.

\begin{table}
\caption{Near-IR bands observed with NIRI at Gemini-North.}
\tiny
\label{obs}
\begin{tabular}{lccccc}
\hline
\hline
\noalign{\smallskip}
Filter      & $\lambda_c $ & Width   & Eff. Resol. & Frames   & Exp. Time  \\
            & ($\mu$m)      & ($\mu$m) & (arcsec)	       &          & (sec)      \\
\hline
\noalign{\smallskip}
\multicolumn{6}{c}{ {\it Broad-bands}  }\\
\hline
\noalign{\smallskip}
{\it J}           &   1.25       &   0.18  &  0.37     &  19      &  343.90    \\
\noalign{\smallskip}
{\it H}           &   1.65       &   0.29  &  0.4      &  19      &   68.4     \\ 
\noalign{\smallskip}
{\it Ks}          &   2.15       &   0.35  &  0.34     &  14      &   15.4 \\
\hline
\noalign{\smallskip}
\multicolumn{6}{c}{ {\it Narrow-bands}  }\\
\hline
\noalign{\smallskip}
{\it H}-cont     & 1.570        &  0.0236 &  0.42     &  44      & 1764.40    \\
\noalign{\smallskip}
[FeII]      & 1.644        &  0.0387 &  0.38     &  44      & 1764.40    \\  
\noalign{\smallskip}
{\it K}-cont     & 2.0975       &  0.0275 &  0.42     &  16      &  129.60    \\  
\noalign{\smallskip}
H$_{2}$ 1-0 S(1) & 2.2139       &  0.0261 &  0.44     &  27      &  218.70    \\
\noalign{\smallskip}
Br$\gamma$   & 2.1686       &  0.0295 &  0.40     &  24      &  194.40    \\
\noalign{\smallskip}
H$_{2}$ 2-1 S(1) & 2.2465       &  0.0301 &  0.38     &  27      &  218.70    \\
\noalign{\smallskip}
CO 2-0 (bh) & 2.289	   &  0.0279 &  0.36     &  29      &  234.90    \\
\hline
\end{tabular}
\end{table}

\section{Results}
\label{res}

\begin{figure}[h!]
\centering
\includegraphics[width=8.5cm]{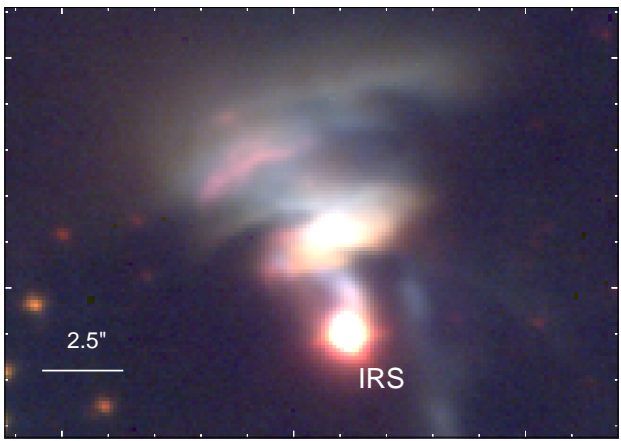}
\caption{Three-colour image with the {\it JHKs} broad-bands emission presented in blue, green, and red, respectively.}
\label{coljhk}
\end{figure}

Figure\,\ref{jhkbroad} displays, in three panels, the emission of the {\it JHKs} broad-bands obtained towards IRS. 
It can be appreciated, with great detail, the presence of a nebula extending to the north/northeast of the point source,
which appears to be attached to IRS by a jet-like structure. The nebula is mainly composed by two arc-like features, the closer
to the source is more intense than the farther one, which is more extended and diffuse. 
In the three bands the nebula connected with the jet-like structure is resolved in a twisted-shaped feature. Besides, 
the emission at the {\it Ks}-band presents a noteworthy bow shock-like structure to the northeast.
Figure\,\ref{coljhk} displays in a three-colour image the {\it JHKs} bands where the mentioned structures on the
three bands can be appreciated  superimposed. It is worth noting that these images
improves, in resolution and sensitivity, the {\it JHKs} images from UKIDSS presented in Paper\,I,
being one of the best near-IR images set obtained towards this kind of nebulosities related to intermediate/high-mass 
YSOs presented up to date.

\begin{figure}[h!]
\centering
\includegraphics[width=7.5cm]{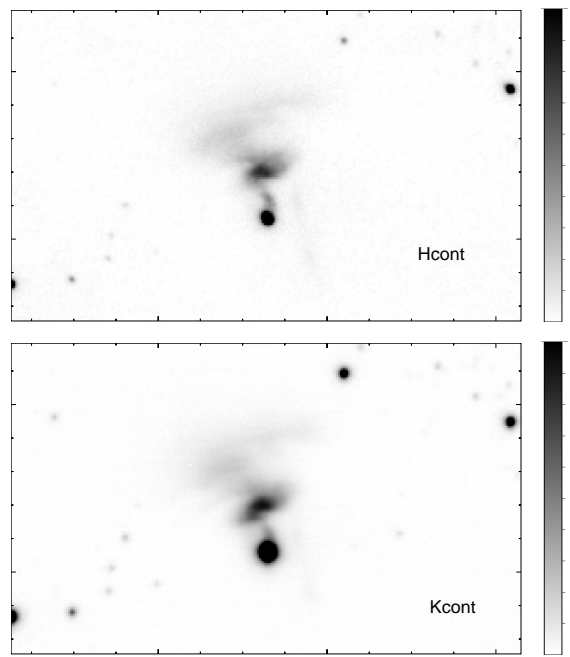}
\caption{Narrow-bands {\it H}- and {\it K}-continuum in upper and bottom panels, respectively. The white and black in the 
colorbars represent 0\% and 100\% of emission, and the maximum values are 3 and 30 ADU for the {\it H}- and {\it K}-cont emission, 
respectively. All images were normalized to 1 sec. }
\label{conts}
\end{figure}

\begin{figure*}[ht!]
\centering
\includegraphics[width=14.5cm]{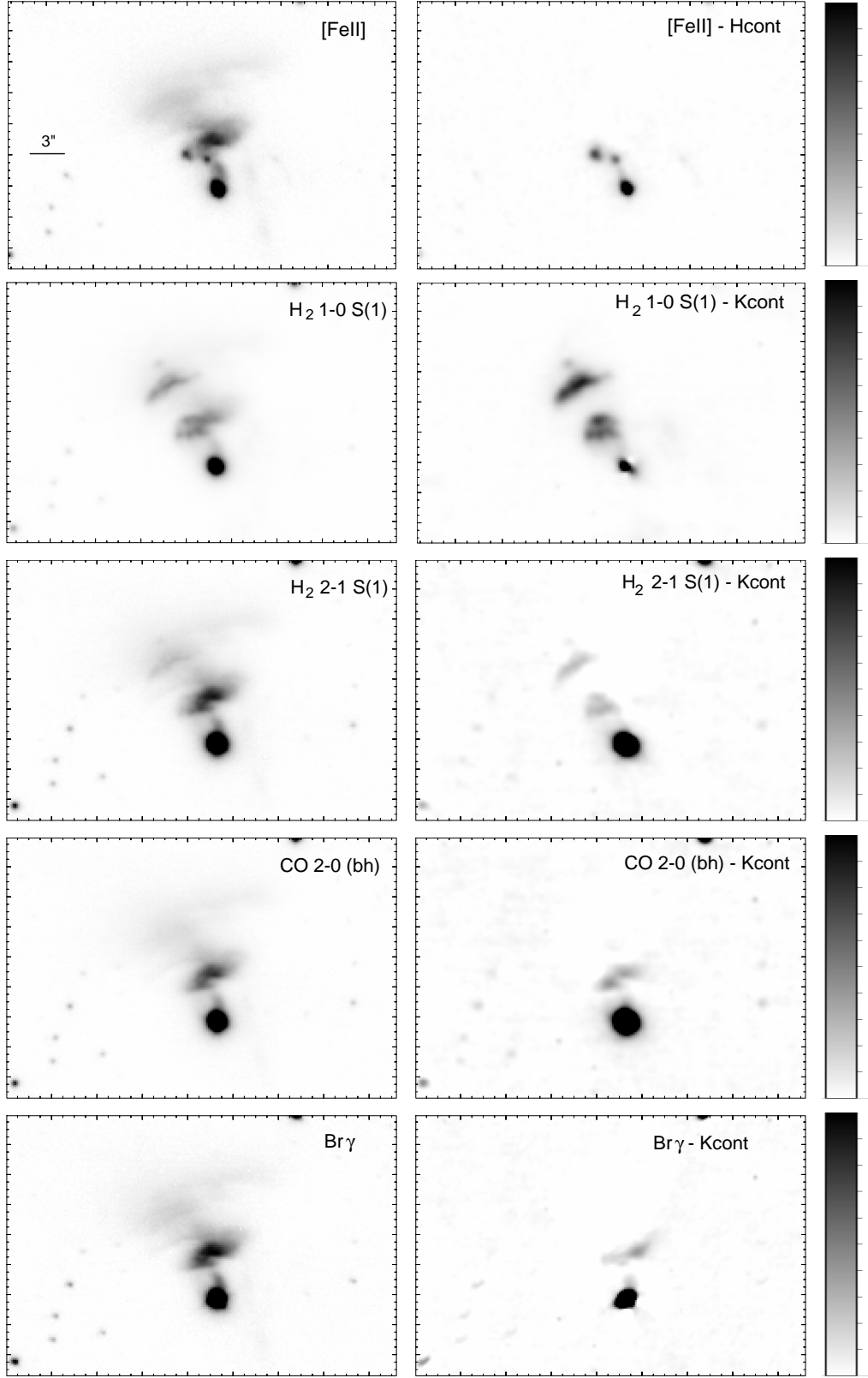}
\caption{Left panels: Narrow-band images centered at the indicated lines with continuum. Right panels: Narrow-band images centered at the indicated 
lines where the corresponding continuum was subtracted. The object nebular
emission in all the subtracted images is above 5$\sigma$ over
the background. The white and black in the 
colorbars represent 0\% and 100\% of emission. The maximum values are (from top to bottom panel): 3, 35, 10, 15, 
and 10 ADU. All images were normalized to 1 sec.}
\label{figsall}
\end{figure*}

Figure\,\ref{conts} presents the continuum emission at the observed narrow-bands {\it H} and {\it Ks}, while
Fig.\,\ref{figsall} displays the narrow-band images, centered at the emission lines of [FeII], H$_{2}$ 1-0 S(1), H$_{2}$ 2-1 S(1), 
Br$\gamma$, and CO 2-0 (bh), with and without continuum (left and right panels, respectively).
As expected, the images of the narrow-band filters with the continuum shows diffuse emission with similar morphology as 
the observed in the broad-band images, whereas some features and knots likely due to the pure line emission are more evident in 
the narrow-band images. Analyzing the continuum subtracted images it can be appreciated that the [FeII] emission shows two bright knots at the 
end of the jet-like structure that extends from the stellar source to the closer arc-like feature. These knots are located at the projected 
distance of 2.8 and 1.7 arcsecs from IRS ($\sim$0.015 and 0.009 pc, respectively at the assumed distance of 1.1 kpc).
In the close surroundings of IRS, the H$_{2}$ 1-0 S(1) line shows several knots located at different distances, most of them lying at the 
arc-like feature closer to IRS (located at 
the projected distance of about 3.2 arcsecs, i.e. $\sim$0.017 pc from IRS), then  
a bow shock feature composed by two arc-like structures at a projected distance of 5.5 arcsecs ($\sim$0.03 pc) from IRS, and 
finally an isolated and farthest knot towards the northeast at a projected distance of 7.4 arcsecs ($\sim$0.04 pc) from the stellar source.
The morphology of the H$_{2}$ 2-1 S(1) emission resembles to that of the lower transition, presenting the same features in the close
surroundings of IRS, but fainter.  
The CO 2-0 (bh) line shows diffuse emission composed by two structures
located at the arc-like feature closer to IRS, although these structures has to be taken with caution as the CO 2-0 (bh) 
emission shape was sensitive to the parameters chosen for the continuum subtraction. 
Finally, the Br$\gamma$ presents also some diffuse emission with a peak slightly shifted with respect the westernmost CO structure.

\begin{figure}[h]
\centering
\includegraphics[width=8cm]{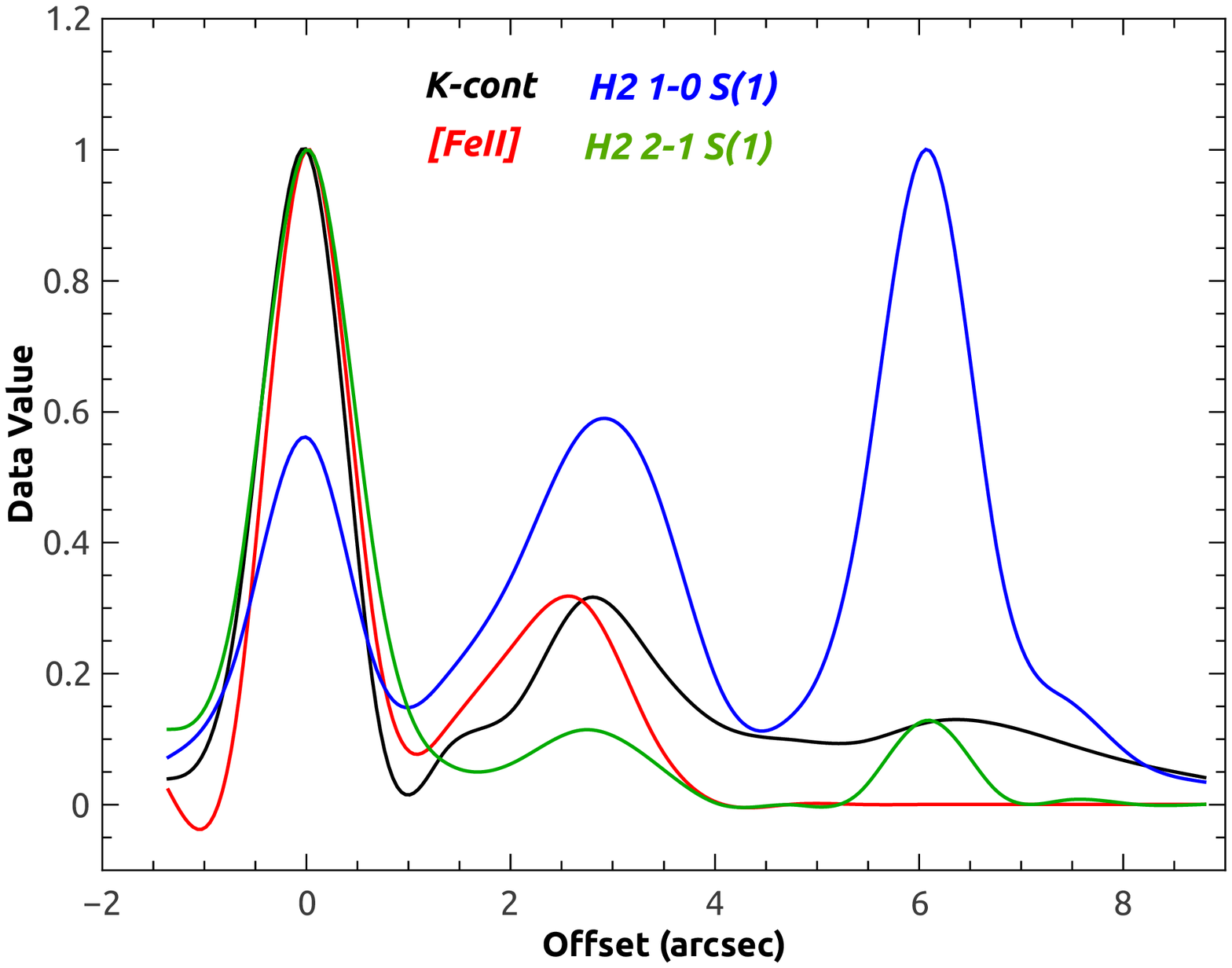}
\includegraphics[width=8cm]{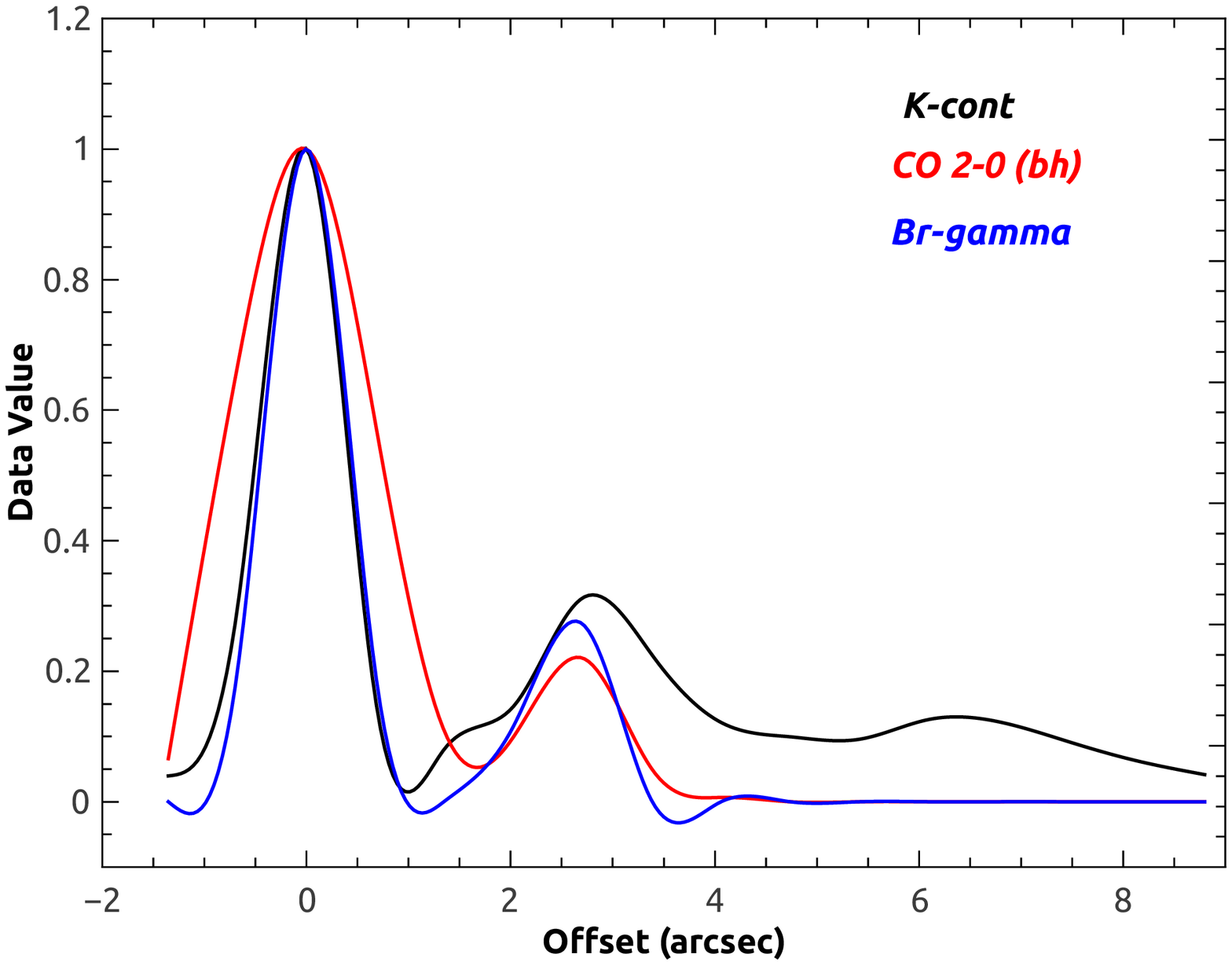}
\caption{Profiles of the line emission plotted along the axis of the cone-like shape nebulosity (the offset axis is
measured in the plane of the sky). In the top panel profiles of [FeII], H$_{2}$ 1-0 S(1), and  H$_{2}$ 2-1 S(1) are
presented in red, blue, and green, respectively. In the bottom panel the profiles of CO 2-0 (bh) and Br$\gamma$ are presented 
in red and blue, respectively. In both panels,
the continuum emission in {\it K}-band, observed with the {\it K}-cont filter (black), is included for comparison. The profiles are 
normalized to the peak emission.}
\label{prof}
\end{figure}

To compare the emission-line features along the cone-like shape nebulosity, we perform a similar analysis as done in \citet{davis11}. 
In Fig.\,\ref{prof} we present the profiles of [FeII], H$_{2}$ 1-0 and 2-1 S(1) (top panel), and CO 2-0 (bh) and
Br$\gamma$ (bottom panel).
In both plots the continuum emission in the {\it K}-band, observed with the {\it K}-cont filter, is included for comparison. 
These profiles were generated by integrating the emissions along a line perpendicular to the cone axis across the whole extension of 
the cone-like shape nebulosity. As the H$_{2}$ 1-0 S(1) extends in knots and filaments much further than the other emission lines 
(see below), for its profile plot we have only included the emission found in the close surroundings of IRS. Table\,\ref{off}
shows the peak positions of each observed feature obtained from Gaussian fits to the profiles presented in Fig.\,\ref{prof}. The errors 
from the Gaussian fits are about 10\%, which in most of the cases are values smaller than the effective spatial resolution of the observations. 

From this analysis it can be appreciated that the emission lines profiles (except the H$_2$ 1-0 S(1)) peak with the highest intensity at 
the position of the IRS source. In similar way, all the emission lines have a second peak between 2 and 3 arcsec offset. There is a third 
peak of emission at about 6 arcsec offset but only for H$_{2}$ 1-0 S(1) and H$_{2}$ 2-1 S(1). These last two emission peaks 
are generated by the two arc-like 
features clearly displayed in the 2-D images. 
An inspection of the 2-D images shows that the second peak 
in the [FeII] profile is generated by projected blend of the profiles of more `compact knots' (in contrast with the more diffuse nebula 
of Br$\gamma$ and H$_{2}$ lines). As mentioned before, both H$_{2}$ emission lines, that trace regions of low-excitation, extend beyond being 
interesting to note here the case of H$_{2}$ 1-0 S(1), for which the projected emission increases with the distance to the source. 
It is important to note that any comparison with the equivalent analysis presented in \citet{davis11} should be done with care because 
both studies are applied to sources of different mass, as in the mentioned work the study is 
on Herbig-Haro type outflow sources.

In order to study the excitation conditions of the H$_{2}$ lines we perform the intensity ratio between the H$_{2}$ S(1) 1-0 and H$_{2}$ S(1) 
2-1 lines towards the three H$_{2}$ structures defined in Fig.\,\ref{prof}, whose offset on the sky from IRS are shown Table\,\ref{off}.
The obtained values are about 10 for the first arc-like feature (first peak of the H$_{2}$ profiles in Fig.\,\ref{prof}), 
and $\sim25$ for the second arc-like feature (second peak in Fig.\,\ref{prof}) and for the 
isolated and farthest knot (third tiny peak in the same figure). In Sec.\,4.1 we discuss these.

\begin{table}
\caption{Offsets of emission-line features along the cone-like shape nebulosity determined from Gaussian fits to profiles shown in Fig.\,\ref{prof}.}
\label{off}
\begin{tabular}{lcc}
\hline
\hline
\noalign{\smallskip}
      & Offset on sky   & Offset on sky  \\
            & (\s)      & (AU)        \\
\hline
\noalign{\smallskip}
[FeII]      & 1.78, 2.56        &  1958, 2816        \\
\noalign{\smallskip}
H$_{2}$ 1-0 S(1) & 2.91, 6.05, 7.66  &  3200, 6655, 8426   \\
\noalign{\smallskip}
Br$\gamma$   & 2.63       &  2893     \\
\noalign{\smallskip}
H$_{2}$ 2-1 S(1) & 2.74, 6.10, 7.64       & 3014, 6710, 8404      \\
\noalign{\smallskip}
CO 2-0 (bh) & 2.68        &   2948   \\
\hline
\end{tabular}
\end{table}

\begin{figure}[h!]
\centering
\includegraphics[width=9cm]{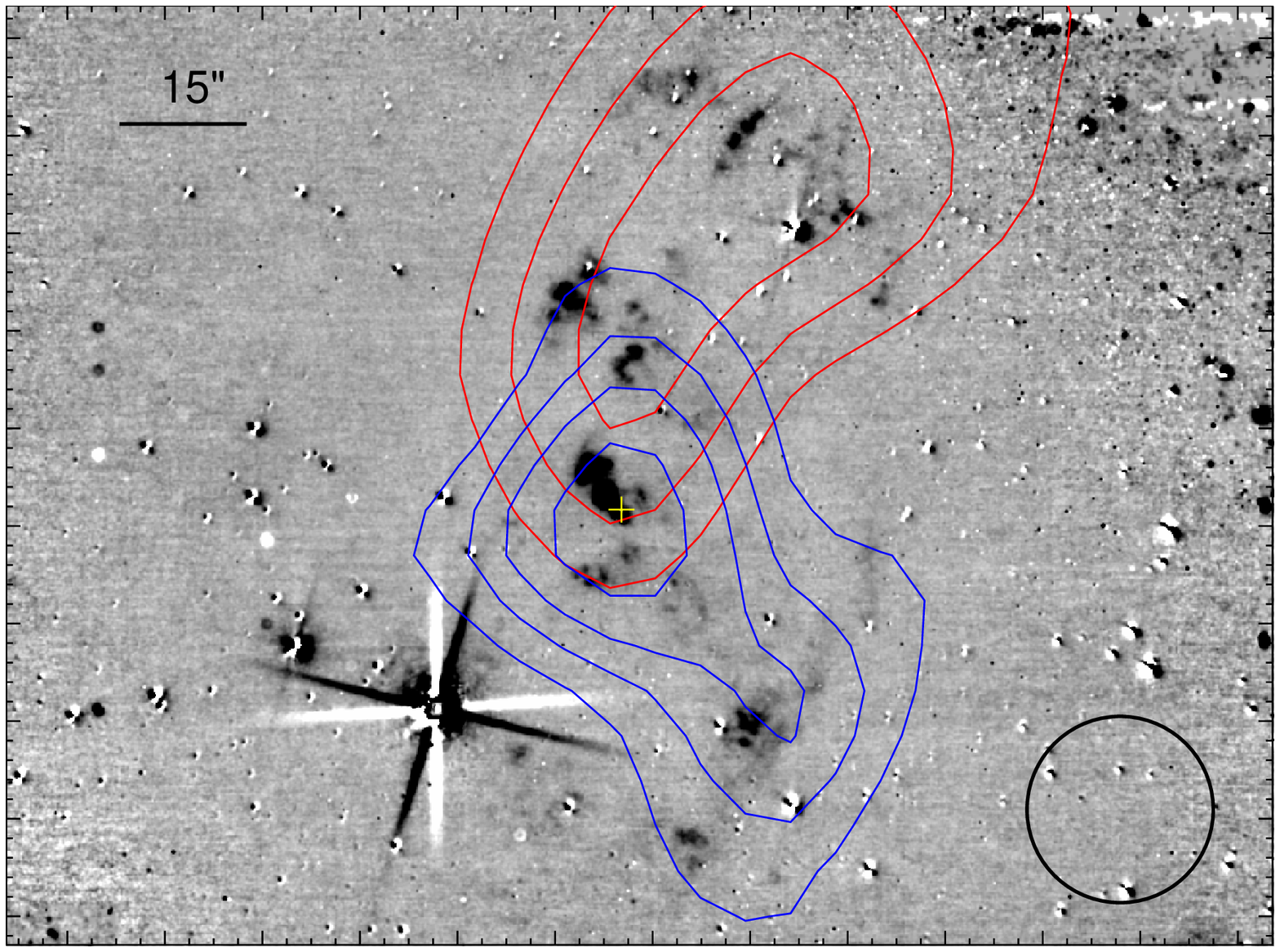}
\caption{H$_{2}$ 1-0 S(1) continuum subtracted
emission with the \2 J=3--2 contours superimposed, as presented in Paper\,I representing the red and blueshifted molecular
outflows. The yellow cross is the IRS position and the beam-size of the \2 observations is shown at the bottom
right corner.}
\label{h2lobes}
\end{figure}

Finally, we have carefully inspected the images looking for emission likely related to IRS at larger spatial scales.
The only case in which emission extends beyond the localized in the IRS close surroundings, is in H$_{2}$ 1-0 S(1).
In the H$_{2}$ 1-0 S(1) continuum subtracted image several knots and filaments
appear at about 45\s~(equivalent to 49.5 kAU in projection) around IRS towards the northwest and southwest, in perfect matching with the 
distribution of the red and blueshifted molecular outflows discovered in Paper\,I. Figure\,\ref{h2lobes} shows the H$_{2}$ 1-0 S(1)
continuum subtracted emission with contours of the integrated \2 J=3--2 line as presented in Paper\,I.

\subsection{Proper motions}
\label{Sprop}

Following the precessing jet scenario and using two observations with temporal difference, it is possible to derive
an estimation of the proper motions of some structure belonging to the cone-like shape nebulosity. To do so we
compare our {\it J}-band observations of the field, taken during August 2014, with the {\it J}-band
image from the UKIDSS Survey taken during June 2005, providing a temporal interval of about 9 years.
The angular resolutions are about 0.1 and 0.4 arcsec pix$^{-1}$~for Gemini and UKIDSS, respectively.
The astrometry of both images was checked based on the position of several point sources in the field
and no evident offset was detected within the resolutions involved.

\begin{figure}[h!]
\centering
\includegraphics[width=8cm]{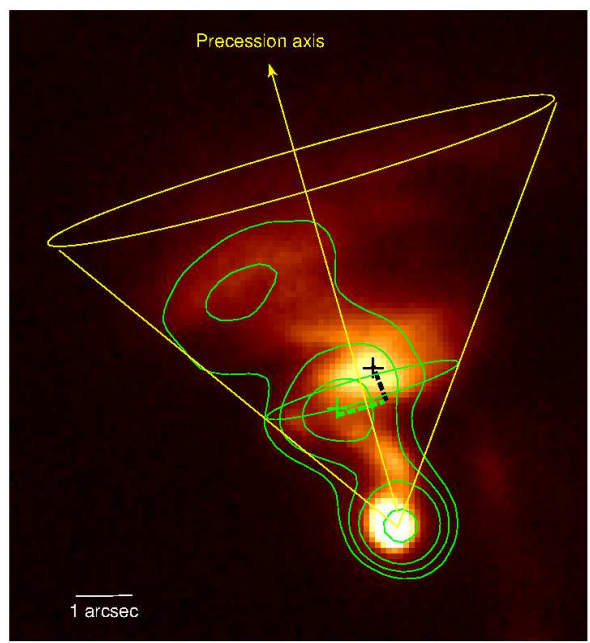}
\caption{Combination of the {\it J}-band emission obtained in two different epochs:
the background image is from our observations obtained in 2014 while the green contours superimposed are from the UKIDSS Survey
image obtained in 2005. The green and black crosses are the peak position
at 2005 and 2014, respectively of the same clump-like feature related to the jet.}
\label{prop}
\end{figure}

Figure\,\ref{prop} shows in the background the {\it J}-band emission obtained from our observations, and superimposed
in green contours, the {\it J}-band emission from the UKIDSS Survey.
Based on the morphology of the emission, we delineate the cone likely
traced by the precessing jet (in yellow). The proper motion of the jet can be roughly characterized based on the
assumption that the jet is moving onto the cone wall. In Fig.\,\ref{prop}, the shift in the emission structure can be
assessed by comparing a clump-like feature, whose peak position is indicated with a green cross for 2005, and with
a black cross for 2014.
Assuming that this feature is moving onto the cone wall, the spatial projected shift of 0.95 arcsec can be decomposed into two components.
One component, perpendicular to the cone axis, related to the jet precession movement, of about 0.80 arcsec (green dashed line)
and the other component, in vertical direction along the cone wall, associated with the ejection movement,
of about 0.51 arcsec (black dashed line). Considering that the clump-like feature has taken
about 9 years to cover the distance represented by the green dashed line, we estimate that it would take about 150~yr to cover
the perimeter of the green circumference shown in Fig.\,\ref{prop} (precession period).
Considering the other component of the proper motion represented by the black dashed line,
we estimate an ejection velocity of about 200 km~s$^{-1}$ for the jet (a lower limit given possible projection effects not considered
here), which is in good agreement with typical jets axial speeds from the literature (100--400 km~s$^{-1}$) \citep{mundt90,konigl00}.
Using these parameters and following \citet{weigelt06} and \citet{smith05}, we can roughly analyze whether the jet precession period is
slow or fast.
In order to do that, it is necessary to compare the jet precession period with the outflow expansion time. This parameter is proportional
to the jet dynamical time $t_{j} = r_{j}/v_{j}$ and the ratio of jet to ambient density, $\eta = \rho_{j}/\rho_{a}$, being
$r_{j}$ and $v_{j}$
the initial jet radius and speed. Additionally, it is inversely proportional to sin($\theta$), where $\theta$ is the half-angle of
the precession cone. Using $v_{j} = 200$ \ks~and $\theta = $35\d, and assuming $r_{j} = 1.7 \times 10^{15}$ cm and $\eta = 10$
\citep{smith05}, we obtain an outflow expansion timescale of about 50 yr. Thus, the fact that the jet precession period
was estimated in 150 yr, three times larger than the outflow expansion timescale, suggests the presence of a slow-precessing jet.
As \citet{smith05} point out, the slow-precessing jets generate helical flows, which is in agreement with the observed near-IR
emission morphology.

\section{Discussion}

The cone-like shape nebulosity observed in the {\it JHKs} broad-bands points to the northeast of IRS opening in a 
wide-angle of about 70$^{\circ}$. This near-IR emission likely arises from a cavity cleared in the circumstellar material and 
may be due to a combination of different emitting processes:
radiation from the central protostar that is scattered at the inner walls of the
cavity, emission from warm dust, and line emission from [FeII] and shock-excited H$_{2}$, among other 
emission lines (e.g. \citealt{reip00,bik05,bik06}). Indeed, as shown in Fig.\,\ref{figsall}, some of these lines were detected.
Additionally, the curved morphology of the jet-like feature and the twisted-shaped nebula, mainly observed in the {\it J}-band, 
strongly suggests the presence of a precessing jet. Models presented in \citet{smith05}, pointed
out that the dominant structure produced by a precessing jet is an inward-facing cone, and particularly, a slow-precessing jet 
leads to helical flows, generating a spiral shaped nebula. From our analysis of proper motions (see Sect.\,\ref{Sprop}),
we not only prove the existence of a precessing jet driven by IRS, but also that the precessing movement is slow 
(precession period $= 150$ yr), giving strong observational support to the numerical models of \citet{smith05}.

\subsection{Emission-line features}

\begin{figure}[h]
\centering
\includegraphics[width=8cm]{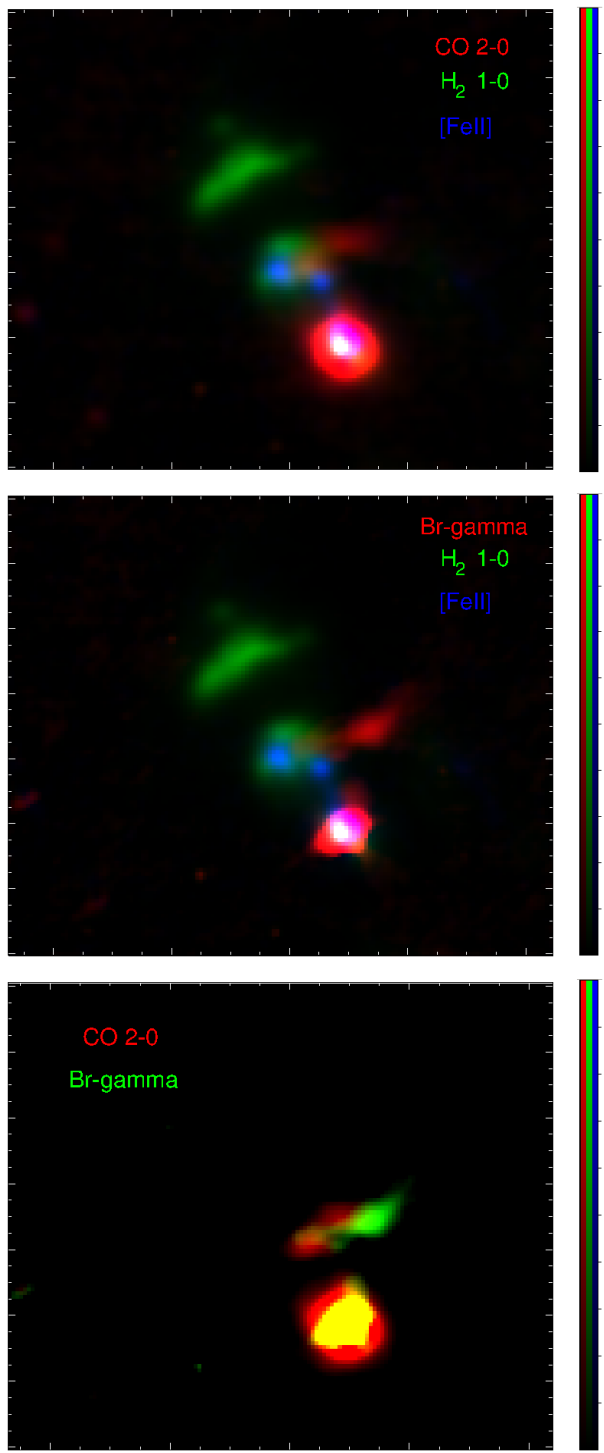}
\caption{Colour images composed with the continuum subtracted lines indicated in each panel.
The black and the respective color in the colorbars represent 0\% and 100\% of emission.
The maximum values are: 8, 50, and 2 ADU for red, green, blue, respectively in top and middle panels,
and 5 ADU for red and green in the bottom panel. All images were normalized to 1 sec.}
\label{colors}
\end{figure}

For comparison, the observed emission lines are presented in colour-composed images
in Fig.\,\ref{colors}, which displays in the top panel, the continuum subtracted CO 2-0 (bh),  H$_{2}$ 1-0 S(1), and [FeII], in the
middle panel the continuum subtracted Br$\gamma$, H$_{2}$ 1-0 S(1), and [FeII], and in the bottom panel the continuum subtracted CO 2-0 (bh) and
Br$\gamma$.
It is known that the [FeII] emission traces the innermost part of jets that are accelerated
near the driving source \citep{reip00}. The [FeII] is closely related to emission knots and shock fronts along the jet
axis, tracing a high-velocity ($v \sim 50-200$ \ks), hot (T $\sim 10^4$ K), dense (electron densities $\sim 10^{5}$ cm$^{-3}$), 
and partially-ionized region \citep{bally07,davis11}.
The H$_{2}$ emission around YSOs usually delineates a slow ($v \sim 10-30$ \ks), low-excitation, shocked molecular gas 
component (T $\sim 2 \times 10^{3}$ K, $n \geq 10^{3}$ cm$^{-3}$) \citep{bally07,davis11}. Besides of the collisional excitation mechanism, 
there is another possible mechanism to excite the H$_{2}$ emission that 
should be considered: the UV fluorescence. One way to discriminate between the collisional and radiative mechanisms is through 
the H$_{2}$ S(1) 1-0/2-1 ratio. In the collisional case, high ratios (about 10) are expected, while the radiative mechanism
produces lower ratios (about 2) (e.g. \citealt{black87,wolfire91,smith95}).
The H$_{2}$ S(1) 1-0/2-1 ratio analysis towards the features observed in the close surroundings of IRS 
shows that they are indeed produced by shocked gas, i.e. the H$_{2}$ emission is collisionally excited. Moreover, taking into 
account that in the closer arc-like feature the [FeII] and H$_{2}$ emissions are spatially coincident (see Fig.\,\ref{colors}, upper and 
middle panels), we suggest that this feature may be produced by a J-shock, in which the [FeII] emission arises in regions where
the shock velocity could be larger than 30 \ks, while the H$_{2}$ emission is originated in regions where the shock velocity is less than 
25 \ks~and the molecule is not dissociated \citep{hollen89,smith94,reip00}, or in the post-shock cooling region, where the molecules
reform and radiate. Thus, even though a J-shock can be dissociative, emission from H$_{2}$ lines
may be detected. In addition, it is important to keep in mind that iron can also be excited in C-shocks \citep{dionatos13,dionatos14}.
On the other side, regions with H$_{2}$ emission without
[FeII] can be explained by both, a nondisocciative C-shock, or a less energetic J-shock \citep{hollen89,dionatos13,dionatos14}.  
This is the case of the second arc-like feature in the surroundings of
IRS and the H$_{2}$ 1-0 S(1) features observed in a larger spatial scale shown in Fig.\,\ref{h2lobes}. As observed in the figure, 
these features perfectly correlate with the \2 outflows presented in Paper\,I, showing the presence of shocked molecular gas along them, and
confirming its nature of molecular outflows.

The CO bandheads are excited in hot and very dense regions (T $>$ 2000 K and n $> 10^{10}$ cm$^{-3}$), 
physical conditions that can be found in accretion disks \citep{sco83,carr89,bik04}, neutral winds \citep{carr89}, and funnel flows 
between the disk and the central source \citep{martin97}. 
There is an intriguing issue regarding the Br$\gamma$ emission around IRS: on one side, Br$\gamma$ is excited at very high temperatures 
(T $\sim$ 10$^{4}$ K) and its emission usually spatially correlates with that of the CO 2-0 (bh) in YSO environments (\citealt{ilee14} 
and references therein), which is not the case for IRS (see
Fig. 8 bottom panel). Besides, according to \citet{kumar03}, the
Br$\gamma$ emission can arise in fast J-shocks within an envelope of
thousands of AU surrounding the young star. If it would the
case for IRS, then it is expected that the Br$\gamma$ emission correlates with the [FeII] emission, 
which is also not the case for IRS (see Fig.\,\ref{colors} middle panel).
We suggest that it is probable that the Br$\gamma$ emission is associated
with stellar wind \citep{kraus08}.

\subsection{Considerations about the orientation of the observed circumstellar structures}
\label{consid}

We dedicate this section to discuss some aspects about the puzzling orientation of the molecular outflows detected in 
\2 J=3–2 and H$_{2}$ 1-0 S(1) lines, compared with that of the jet, the arclike features, and knots observed in the near-IR lines.

As shown in Paper\,I and reinforced in this study through the H$_{2}$ 1-0 S(1) extended emission, which maps shocked molecular gas, 
the molecular outflows are indeed highly misaligned. The redshifted outflow points towards the northwest, while the blueshifted outflow points
towards the southwest. 
Concerning the near-IR features in the close environment of IRS, as observed towards other similar sources 
(see \citealt{prei03,massi04,kraus2006,weigelt06,paron13}) 
the nebula, with a cone shape and arc-like
features, extends only to one side. In the mentioned studies, 
to justify this unidirectional asymmetry, it was proposed that 
the observed near-IR features should be related to a blueshifted 
jet with the redshifted counterpart not detected in the near-IR 
bands due to higher extinction. Indeed, when molecular outflows 
were observed through molecular rotational transitions, the nebulosity 
observed in the near-IR images matches the orientation 
and alignment of blueshifted molecular outflows.
In the case presented here, the fact that the blueshifted molecular outflow points towards the southwest and the 
observed cone shape nebula with arc-like features towards the northeast is indeed puzzling.  
It may suggest that they are not related, or that probably there exist a complex relation, 
not evidenced in these images due to the large difference in the angular resolutions between the millimeter and near-IR data.
One possibility could be that we are observing a not resolved system of YSOs. At the assumed distance of 1.1 kpc, 
the spatial resolution presented here is about 400 AU. It is known that a region of this size, or even smaller, 
can contain a binary system \citep{conn08}. 
In an scenario with more than one YSO, it can be expected a complex (and so far unusual when comparing with low-mass YSO outflows studies) 
distribution of jets and molecular outflows, as it could be the case for IRS.
In this way, the existence of a precessing jet may be explained through tidal interactions between companion stars.

Additionally, by inspecting the \2 J=3--2 distribution, it can be noticed that the peak of the blueshifted component coincides in projection
with IRS. This shows that the \2 J=3--2 emission around IRS is predominantly blueshifted, which would relate the near-IR cone shape nebula to 
blueshifted gas. This agrees with an scenario with more than one YSO, where the blueshifted \2 J=3--2 peaking at IRS and the lobe extending 
towards the southwest are produced by different sources. Interferometric rotational lines CO observations are required to resolve 
the molecular outflows morphology.

\section{Summary and concluding remarks}

In Paper\,I we reported the presence of misaligned molecular outflows
towards the intermediate-mass YSO UGPSJ185808.46+010041.8 (IRS) and based
on public near-IR data we suggested the presence of a precessing jet. 
Aiming at studying in more detail this interesting source we present here the results derived from a new high-resolution 
image set obtained with Gemini-NIRI in the JHKs broad-bands and [FeII], H$_{2}$ 1-0 (S1), Br$\gamma$, 
H$_{2}$ 2-1 (S1), and CO 2-0 (bh) narrow-bands.

The near-IR imaging towards IRS, obtained with an spatial resolution of about 400 AU at the assumed
distance of 1.1 kpc, strongly suggest a precessing jet. The observed cone-like shape nebula 
composed by a twisted-shaped feature with two arc-like structures is one of the best images of this 
kind of objects presented up to date. These images give an important observational support to the models 
that point out that precessing jets generate this kind of near-IR features in massive YSOs.
An analysis of proper motions based on our Gemini observations and UKIDSS data gives additional support to the
precession scenario and allowed us to estimate a precession period of 150 yr, a jet axial speed of about 200 km~$^{-1}$,
and an outflow expansion timescale of about 50 yr. These parameters suggest the presence of a slow-precessing jet, 
which as shown by models presented in \citet{smith05}, leads to helical flows in agreement with the structures  
observed in our near-IR images.

The analysis of the observed near-IR lines shows that the H$_{2}$ is collisionally excited, 
and the spatially coincide of the [FeII] and H$_{2}$ emissions in the closer arc-like feature 
suggests that this region can be affected by a J-shock. However, it is important to keep in mind that iron can also be excited in C-shocks. 
The second arc-like feature presents H$_{2}$ emission without
[FeII], which could be explained by nondisocciative C-shock or a less energetic J-shock, 
as is the case of the H$_{2}$ 1-0 S(1) features observed in a larger 
spatial scale. In the H$_{2}$ 1-0 S(1) continuum subtracted image, the only case in which near-IR emission was detected extending 
beyond the localized 
in the IRS close surroundings, several knots and filaments
appear in perfect matching with the distribution of the molecular outflows discovered in Paper\,I, confirming
shocked gas within the outflow lobes.
Finally, we suggest that the Br$\gamma$ emission around IRS probably arises from stellar winds.  

It is really puzzling the orientation of the molecular outflows (pointing the redshifted one towards the northwest and the blueshifted one
towards the southwest), compared with the near-IR features (pointing towards the northeast).
One possibility is a not resolved system of YSOs. 
The \2 J=3--2 emission around IRS is predominantly blueshifted, which suggests a relation between the near-IR cone shape nebula and
the blueshifted gas. In this case, the blueshifted \2 J=3--2 peaking at IRS and the lobe extending
towards the southwest can be produced by different sources.  This scenario can explain the existence of a precessing jet 
through tidal interactions between companion stars. Interferometric rotational lines CO observations are required to resolve
the molecular outflows morphology.

\section*{Acknowledgments}

We thank the anonymous referee for her/his helpful comments and suggestions.
The authors are very grateful to Gemini Staff for their dedication and helpful support in carrying out our observational programme.
S.P. and  M.O. are members of the {\sl Carrera del 
Investigador Cient\'\i fico} of CONICET, Argentina. 
This work was partially supported by Argentina grants awarded by UBA (UBACyT), CONICET and ANPCYT.

\bibliographystyle{aa}  
\bibliography{ref}
\IfFileExists{\jobname.bbl}{}
{\typeout{}
\typeout{****************************************************}
\typeout{****************************************************}
\typeout{** Please run "bibtex \jobname" to optain}
\typeout{** the bibliography and then re-run LaTeX}
\typeout{** twice to fix the references!}
\typeout{****************************************************}
\typeout{****************************************************}
\typeout{}
}

\label{lastpage}
\end{document}